\documentstyle[preprint,eqsecnum,epsf,aps]{revtex}

\def\Journal#1#2#3#4{{#1} {\bf #2}, #3 (#4)}

\def\NPB{{\em Nucl. Phys.} B}
\def\PLB{{\em Phys. Lett.}  B}
\def\PRL{{\em Phys. Rev. Lett.}}
\def\PRD{{\em Phys. Rev.} D}

\def\JHEP{{\em J. High Energy Phys.}}

\def\be{\begin{equation}}
\def\ee{\end{equation}}
\def\bea{\begin{eqnarray}}
\def\eea{\end{eqnarray}}

\def\Lgr{{\cal L}}
\newcommand{\vev}[1]{\langle#1\rangle}
\def\be{\begin{equation}}
\newcommand{\bel}[1]{\begin{equation}\label{#1}}
\def\ee{\end{equation}}

\newcommand{\rem}[1]{}
\def\tr{{\rm tr}}
\def\half{{1\over 2}}

\def\NN{{\cal N}}

\def\none{$\NN=1$}
\def\ntwo{$\NN=2$}

\def\nfour{$\NN=4$}

\def\ZZ{{\bf Z}}

\newcommand{\AmS}{{\protect\the\textfont2
  A\kern-.1667em\lower.5ex\hbox{M}\kern-.125RMS}}

\def\slash{\!\!\!\!/ \ }

\def\NN{{\cal N}}

\def\none{$\NN=1$}
\def\ntwo{$\NN=2$}

\def\nfour{$\NN=4$}

\def\Lgr{{\cal L}}

\def\ZZ{{\bf Z}}

\renewcommand{\thanks}{\footnote}
\def\tocite#1{$^{\hbox{\,-}}$\kern-.04em\cite{#1}}

\def\JLone<#1,#2>{#1}
\def\JLtwo<#1,#2,#3>{#2}
\def\JLyear<#1,#2,#3,#4>{#3}
\def\JLpage<#1,#2,#3,#4>{#4}

\def\Jpage<#1,#2,#3>{#3}

\begin{document}

\title{On methods for extracting exact 
non-perturbative results in non-supersymmetric gauge theories}
\author{Matthew J. Strassler}
\address{
Dept. of Physics and Astronomy,
University of Pennsylvania, \\
209 S. 33rd St.,
Philadelphia, PA 19104, USA }

\date{\today}
\maketitle
\begin{abstract}
At large $N$, a field theory and its orbifolds (given by projecting
out some of its fields) share the same planar graphs. If the parent-orbifold
relation continues even nonperturbatively, then properties such as confinement
and chiral symmetry breaking will appear in both parent and orbifold.  
\none\ supersymmetric Yang-Mills has
many nonsupersymmetric orbifolds which resemble QCD.
A nonperturbative parent-orbifold relation gives
a number of interesting predictions, exactly valid at large $N$,
and expected to suffer only mild $1/N$ corrections.  These include
degeneracies among bosonic hadrons and exact predictions for
domain wall tensions.  Other predictions are valid
even when supersymmetry in the parent is broken.  Since these theories
are QCD-like, simulation is possible, so these
predictions may be numerically tested.  The method also
relates wide classes of nonsupersymmetric theories.
\end{abstract}
\pacs{Valid PACS appear here.
{\tt$\backslash$\string pacs\{\}} should always be input,
even if empty.}

\narrowtext

In recent years, many exact results for four-dimensional
supersymmetric theories have been obtained.  Supersymmetric theories
have holomorphic or otherwise mathematically special sectors which are
strongly constrained.  While the theories are not exactly soluble,
many remarkable nonperturbative predictions can still be made.  These
include the infrared behavior of some theories, the large
gauge-coupling behavior of others, and the properties of some theories
with many colors at large 't Hooft coupling.

Unfortunately, few of the techniques hold for ordinary,
nonsupersymmetric theories.  This is regrettable for many reasons.  In
nonsupersymmetric solutions of the hierarchy problem, nonperturbative
dynamics is sure to play a role, and we are unable to make any
predictions about it.  In supersymmetric theories, supersymmetry
breaking is often a strongly-coupled phenomenon, and again we can make
very few predictions in the strong-coupling sector.  Applications of
quantum field theory to condensed matter (especially in three
dimensional systems) and to statistical mechanics are limited by our
lack of nonperturbative knowledge.  Finally, no lattice simulations of
four-dimensional supersymmetric theories (with one exception discussed
below) are remotely practical, and so the techniques of supersymmetric
theories cannot be tested numerically, nor can additional information
be obtained through simulation.

There is consequently ample motivation for attempting to find new
nonperturbative methods in nonsupersymmetric theories.  So little is
known that any progress whatsoever is of value.  In this letter I
employ a general technique --- orbifolding, explained below --- which
predicts relations {\it between} gauge theories that are exact in the
limit where the number of colors is taken to infinity.  Of course, a
statement that theory A and theory B have the same Green's functions,
while remarkable, does not give direct predictions in either theory A
or theory B alone.  To obtain quantitative predictions in theory B one
must have some additional {\it a priori} knowledge of theory A,
obtained from another source.  In this letter I point out that if we
take ${\cal N}=1$ supersymmetric Yang-Mills theory as theory A, about which
a number of interesting facts are known, some of them exactly, we can
obtain predictions, some of them exact (up to $1/{N}$ corrections
with no exceptionally large coefficients) for a class of
nonsupersymmetric theories.  Furthermore, the relations between these
theories survive even when supersymmetry is broken, for any value of
the supersymmetry-breaking parameter.


The technique used here involves comparing a
``parent'' theory to an ``orbifold'' of the parent
theory.\footnote{ Orbifolding involves imposing a certain
discrete projection operator on the degrees of freedom of a field
theory, but there are two distinct meanings to the term.  
One involves orbifolding
the space-time in which a field theory is defined.  The second, 
referred to in this paper, involves projecting out some of the
fields of a theory \cite{dougmoor,BerKakVaf,berjoh,Schmorbi}, leaving the space-time unaffected.}  The orbifold theory is obtained from the
parent by judiciously setting a number of degrees of freedom to zero
--- that is, by inserting appropriate delta functions of the fields
into the path integral.  (Notice this is not a projection operator on
the generating functional itself, but on the measure of integration
defining the generating functional.)  Following discoveries in the
gauge-gravity correspondence \cite{KachSilver,LawNekVaf} it was pointed
out that there is a striking relationship even at weak coupling
between orbifold theories and their parents \cite{BerKakVaf,berjoh}.
Specifically, if one organizes color indices using 't Hooft's
double-line notation, and uses his topological classification of
Feynman diagrams, then parent and orbifold share the same planar
graphs (up to a trivial rescaling of coupling constants.) This means
the two theories have many Green's functions in common at large $N$,
at least to all orders in perturbation theory.  Note that
supersymmetry is  nowhere needed in this argument.

More precisely, consider the Green's functions
of the parent theory, which are functions of the Yang-Mills coupling
$g^2_{YM}$ and the number of colors $N$.  Define $\zeta \equiv
g^2_{YM} N / 8 \pi^2$, $\alpha = g^2_{YM}/4\pi$; then the Green's
functions may be reexpanded, as 't Hooft suggested, as
$G^{p}(\zeta,\alpha)$; here $p$ stands for ``parent''.  The limit
$\alpha\to 0$ with $\zeta$ fixed keeps only the planar graphs in the
perturbative expansion.  The orbifolds of this theory (under suitable
conditions) will share the same planar graphs, so their Green
functions $G^{o}(\zeta,\alpha)$ will share the same limit;
$G^o(\zeta, 0) = G(\zeta,0)$.  However, the number of fields in
the orbifold is less than the number in the parent; for example, if
the orbifold involves a $\ZZ_k$ symmetry, then the number of fields is
reduced by a factor $k$.  This means that to keep the parent and
orbifold value of $\zeta$ equal, we must set $\alpha^{(o)} =
k\alpha^{(p)}$.  This scaling trivially cancels out of the planar
graphs. 

However, while rigorous, this argument does not imply that orbifold
and parent share their non-perturbative effects, for example effects
of order $e^{-1/\zeta}$.  Unfortunately, such effects include
confinement and chiral symmetry breaking, string tensions, pion masses
and baryon masses, among others --- in short, all of the most
interesting dynamics of QCD.  It is pure conjecture --- let us call it the
``non-perturbative orbifold'' (NPO) conjecture --- that {\it the
parent and orbifold theories have the same non-perturbative properties
at large $N$.} 

As stated, the conjecture is still imprecise and must be sharpened.  
In general the parent and orbifold theories do not have the
same vacua.  The parent has vacua lacking in the orbifold, because
they are projected out by the orbifolding process.  The orbifold has
extra vacua not shared by the parent, ones which break spontaneously
the orbifold symmetry, which appears as a global symmetry group in the
orbifold theory.  Similarly, the parent has operators absent in the
orbifold (since they are projected out) and the orbifold has operators
absent in the parent (which are non-singlets under the orbifold
symmetry group.)  The statements concerning planar graphs are that
{\it in those vacua which appear in both theories, the Green's
functions for operators which appear in both theories are identical,
up to appropriate rescalings of coupling constants.}  The NPO
conjecture simply suggests this extends to nonperturbative effects.

 If the NPO conjecture is true, then it relates the nonperturbative
properties of field theories which have small $1/N$ corrections,
independent of whether they have supersymmetry.  But is there any
evidence in its favor?  In supersymmetric theories, even those with
only \none\ supersymmetry, there is considerable evidence for the NPO
conjecture both from the AdS/CFT correspondence
\cite{KachSilver,LawNekVaf,HSU} and from nonperturbative exact results
in supersymmetric theories \cite{HSU,LykPopTri,ErlNaq}. Unfortunately,
little evidence exists for nonsupersymmetric orbifolds.  In my view,
most of the work on nonsupersymmetric orbifolds in the gauge-gravity
correspondence is suspect.  The examples discussed initially
\cite{KachSilver,BerKakVaf,FraVaf} have massless scalar fields, and
are subject to large $1/N$ corrections (see
Refs. \cite{csakietal,IRstable}) which can completely alter their
nonperturbative properties.  In one case, however, a
nonsupersymmetric theory was discussed which does not have large $1/N$
corrections \cite{jpms}. The parent theory is so-called ``\none$^*$'',
namely \nfour\ Yang-Mills, softly broken to \none\ Yang-Mills by mass
terms.  The orbifold of this theory is a finite \ntwo\ supersymmetric
theory broken softly to a nonsupersymmetric $\ZZ_2$ orbifold of \none\
Yang-Mills.  There are no large $1/N$
corrections in this theory, as there are no unprotected light scalar
fields, continuous moduli spaces, or nearly marginal operators which
are unstable to $1/N$ corrections.  It was found \cite{jpms} that the
parent supersymmetric theory and its nonsupersymmetric orbifold share
the same physics, including confinement, monopole/dyon condensates,
flux tubes with appropriate quantum numbers, and so forth, and that
the expected quantitative relations between the theories do hold.
Thus the orbifold conjecture does apply for at least one
QCD-like nonsupersymmetric theory, up to controllably small $1/N$ corrections,
and in particular reproduces interesting nonperturbative phenomena at
large $\zeta$.  This motivates us to take the conjecture seriously,
even without proof, and to see where it leads us.

I will now discuss some predictions of the NPO conjecture, 
and then provide some circumstantial evidence that
it is likely to be correct in this context.  Let us take, as parent
theory, \none\ $SU(N)$ supersymmetric Yang-Mills theory, which
consists simply of a gauge boson and a fermion (``gaugino'') in the
adjoint representation, with Lagrangian
$$
\Lgr = {1\over g^2}\tr \ (-\half F^2 + i\bar\lambda D\slash \lambda ) \ .
$$
Much is known about this theory.  It has an anomalous $U(1)$ axial
symmetry, of which a $\ZZ_{2N}$ subgroup is anomaly-free. It generates
a gluino condensate $\vev{\lambda\lambda}$, which breaks the
$\ZZ_{2N}$ to $\ZZ_2$.  This condensate, of size $ \vev{\lambda
\lambda}\propto e^{-1/\zeta} $, does not appear in the planar graphs,
yet is not suppressed at large $N$.  The phase of the gluino
condensate (let us call it the $\eta'$, in analogy with the similar
phase in QCD) has a potential with $N$ isolated vacua, and in each the
$\eta'$ has a mass of order $e^{-1/\zeta}$.  The domain walls between
these vacua have tensions of order $\vev{\lambda\lambda}$.  The theory
confines and has flux tubes carrying a $\ZZ_N$ charge, with string
tensions and hadron masses of order $e^{-1/\zeta}$.  Moreover the
strings can annihilate in bunches of $N$.  In short, almost all the
interesting properties of the theory are absent in the pure planar
graphs, and many depend on $\alpha$ or $N$.  This is why the orbifold
conjecture becomes especially interesting if it applies
nonperturbatively.

We can apply the conjecture to simple $\ZZ_p$ orbifolds of this theory
which are nonsupersymmetric and similar to QCD.\footnote{To my
knowledge these theories first appear in \cite{georgi,kaplan}, were  
viewed as orbifolds only recently \cite{Schmorbi}, and were
emphasized as especially interesting first in \cite{jpms,ioffeQCD}.}  
A clear field-theorist's
introduction to these theories has appeared in Ref.~\cite{Schmorbi}
and the reader is directed there for further discussion.  For the
string theorist, the $\ZZ_2$ orbifold in question is discussed in
Ref.~\cite{jpms}.  For our purposes here, I will simply
state the results of \cite{Schmorbi}.

The $\ZZ_p$ orbifold of $SU(pN)$ SYM theory is simply given
by breaking the $pN\times pN$ matrices into $N\times N$ blocks, keeping
the diagonal blocks for the gauge bosons, and keeping the
just-above-the-diagonal blocks for the fermions.  The resulting theory
is $SU(N)^p$ --- let us call the $p$ factors $SU(N)_1$, $SU(N)_2$,
etc. --- with $p$ Weyl fermions $\psi_i$, $i=1,2,\dots,p$, charged as
${\bf N}$ under $SU(N)_i$ and as ${\bf\overline N}$ under
$SU(N)_{i+1}$ (or for $\psi_p$ as ${\bf\overline N}$ of $SU(N)_1$.)
``Moose'' or ``quiver'' diagrams of the theories with $p=1,2,3,4$ are
shown in the Figure.\footnote{Strictly speaking, we ought to consider
$U(pN)$ and its orbifolds $U(N)^p$.  However, the missing $U(1)$
factors (except one, for $p$ even) are anomalous and presumably are
massive.  This subtlety is a $1/N$ effect, of course.}

\newpage
\begin{figure}
\centering
\epsfxsize=5.0in
\hspace*{0in}\vspace*{0in}
\epsffile{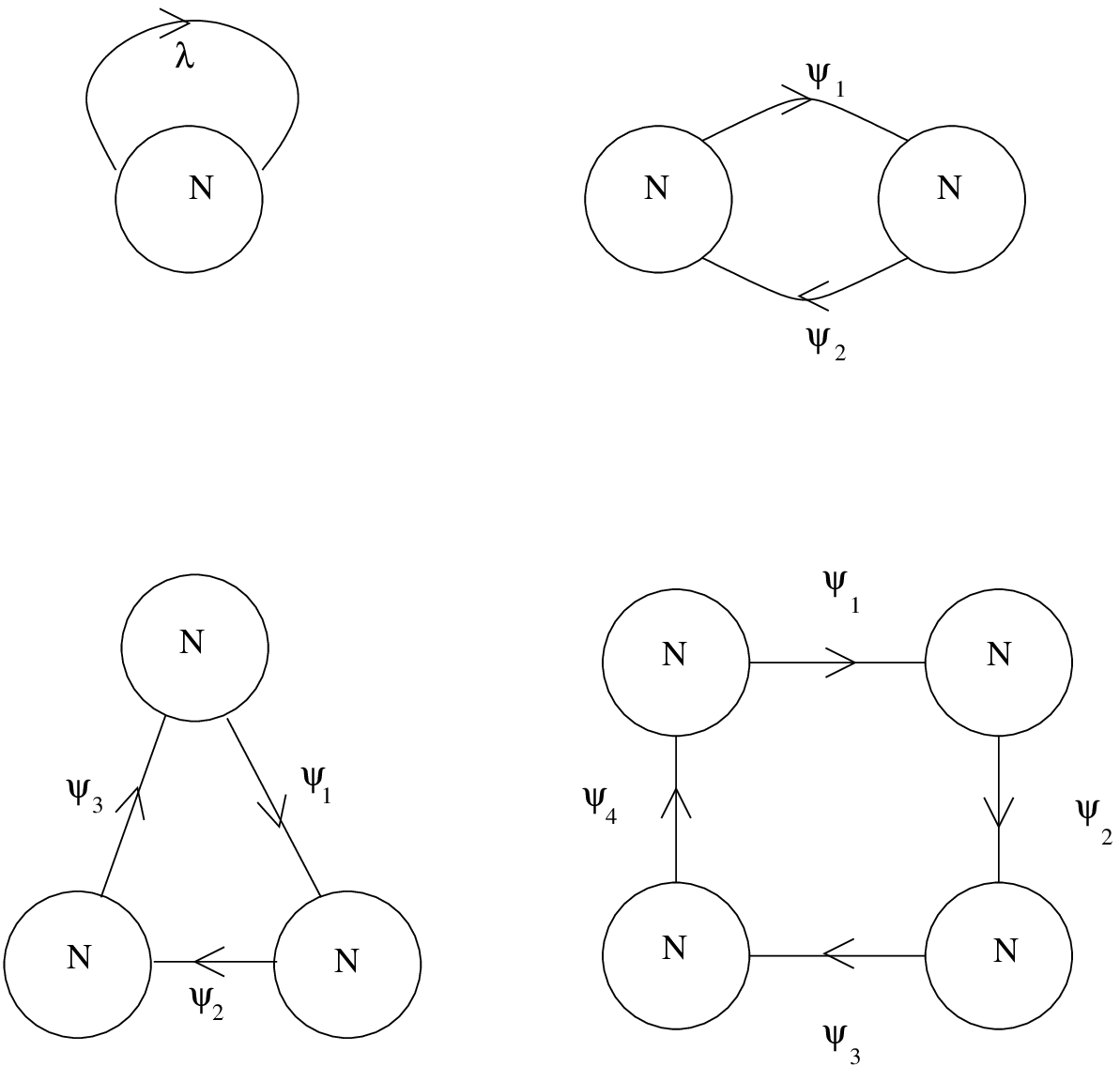}
\caption{\none\ supersymmetric Yang-Mills and its
$\ZZ_2$, $\ZZ_3$ and $\ZZ_4$ orbifolds.}
\label{fig:walls}
\end{figure}
\newpage

Let us first discuss the symmetries of these theories.  As always, the
$\ZZ_p$ symmetry used to orbifold the SYM theory is a symmetry
$(\ZZ_p)_O$ of the resulting orbifold gauge theory; for each
$q=1,\dots,p$, it simply cyclically shifts the $i^{th}$ gauge group
and fermion into the $(i+q)^{th}$ group and fermion.  There is also a
reflection symmetry $i \to p-i+1$.
Also, each one of the $SU(N)$ gauge groups has a $\ZZ_N$ center.  The
matter in the theory is invariant under a simultaneous gauge
transformation, in each of the $SU(N)$ factors, by $U = \omega {\bf
1}$, where $\omega$ is any $N^{th}$ root of unity. This symmetry I
will call $(\ZZ_N)_C$, where $C$ stands for ``center of the
gauge group''. This group determines the classes of electric sources
which can be used to probe the theory and the symmetry group of any flux
tubes should the theory be confining.

There are also classical $U(1)$ symmetries rotating each of the $p$
fermions, but these are anomalous.  Let us consider rotations of the
$\psi_i$ by $e^{2\pi\sigma_i}$.  To be anomaly free, a rotation must
have $N(\sigma_i-\sigma_{i-1})$ an integer.  Accounting for the
centers of the $SU(N)_j$ groups, we have the transformation
$\psi_1\to\psi_1\exp(2\pi i /N)$ with all other $\psi_i$ invariant.  I
will refer to this as $(\ZZ_{N})_A$, ``$A$'' for axial.  For $p$ odd,
this is expanded to a $(\ZZ_{2N})_A$ by the addition of the rotation
$\psi_j \to \psi_j\exp(-i\pi [-1]^j/N)$; the square of this
transformation is equivalent, using the centers of the $SU(N)$ groups,
to the transformation $\psi_1\to\psi_1\exp(2\pi i /N)$ with all other
$\psi_i$ invariant.  For $p$ even, we have instead an additional
$U(1)_B$, a baryonic symmetry under which $\psi_j\to \psi_j \exp(-i
[-1]^j\alpha)$, where $\alpha$ is an arbitrary phase.  This is a 
first hint that the odd and even $p$
cases are distinct in important ways.

For both odd and even $p$, we might expect that the discrete axial
symmetry will be broken by a fermion condensate, preserving a $\ZZ_2$
$\psi_i\to -\psi_i$ for $p$ odd, and preserving $U(1)_B$ for $p$ even.
If this happens, the resulting theory in both cases will have $N$
degenerate vacua, just as in \none\ SYM.  We might also expect
confinement, with flux tubes charged under $(\ZZ_N)_C$.  We will
see this is plausibly the case.

Let us now explore the $\ZZ_2$ orbifold in detail, and learn what the
NPO conjecture predicts for it.  This theory is especially
interesting, as it resembles QCD.  It contains $SU(N)\times SU(N)$
with a {\it Dirac} fermion $\Psi = (\psi_1,\bar\psi_2)^T$ in the
(${\bf N},{\bf \overline N}$) representation.  Note that the
masslessness of the fermion is protected by the $(\ZZ_N)_A$ discrete
axial symmetry.  For $N=3$, this is just QCD with three massless
quarks, with the diagonal subgroup of the chiral $SU(3)_L\times
SU(3)_R$ flavor symmetry gauged.  Thus, we expect this theory to have
confinement and chiral symmetry breaking also, at least when one of
the couplings is taken very small.  Specifically, if $SU(N)_1$ becomes
strongly coupled, it should generate a condensate for $\psi_1\psi_2$.
Were $g_2=0$, there would be $N^2-1$ pions, along with the $\eta'$,
just as in QCD.  However, once $SU(N)_2$ is gauged, $SU(N)_L\times
SU(N)_R$ of flavor is explicitly broken; the pions become massive,
with masses of order $g_2\Lambda_1$.  With the continuous space of
vacua spanned by the pions removed by their masses, the remaining
vacua of the theory are set by the minima of the $\eta'$ potential
energy.  Because of the anomaly, this potential is a periodic function
with $N$ minima, related by the $(\ZZ_N)_A$ axial symmetry. We thus
find, in this QCD-like theory with weakly gauged flavor, that there
are indeed $N$ degenerate and isolated vacua from the breaking of
$(\ZZ_N)_A$.  With this structure, we expect stable domain walls in
the theory.  Thus $SU(N)_1$ generates chiral symmetry breaking and
domain walls.  But it does not generate flux tubes; any putative
flux tubes can break through pair production of $\psi_1$ and $\psi_2$,
just as flux tubes in QCD break by quark-antiquark pair production.

However, there is no massive matter charged under $SU(N)_2$ (since the
pions are massive) so if $g_2$ is small, the low-energy theory is a
pure Yang-Mills theory of $SU(N)_2$.  This theory confines and makes
flux tubes.  The $(\ZZ_N)_C$ symmetry, which is absent if $g_2=0$,
ensures there are flux tubes that carry a $\ZZ_N$ charge and
do not break.  For example, sources in the (${\bf N},{\bf 1}$)
representation of the gauge group cannot be screened by pair
production of $\psi_1$ or $\psi_2$, so they will be confined by
a flux tube.

Thus, for $g_1 > g_2$, we find chiral symmetry breaking at a large
scale, and flux tubes with tensions at a much lower scale.  What
should happen at the orbifold point, where the couplings are equal?
This is far from clear, since it might be that certain effects due to
$SU(N)_2$ cancel effects produced by $SU(N)_1$.  However, if we assume
that there are no phase transitions as $g_2$ becomes of order $g_1$,
the theory will maintain its confinement, its bilinear fermion
condensate and its $N$ vacua.  This would mean that its nonperturbative
properties would be at least qualitatively similar to those of the parent
\none\ SYM theory, leading us to hope that the NPO conjecture is
correct.  (As mentioned earlier, there is additional evidence in
this case from the gauge-gravity correspondence \cite{jpms}.)  

So let us suppose that it is right; what will happen in this QCD-like
theory?  In the limit $g_1=g_2$ there is the $(\ZZ_2)_O$ symmetry
exchanging the two gauge groups.  All states in the theory will be
even or odd under this $\ZZ_2$, and the NPO conjecture applies to
those which are even.  Let us consider the $\ZZ_2$-even hadrons.  By
gauge invariance, all mesons in this theory (that is, all particles
with less than $N$ constituents) will be bosons.  This is in contrast
to the parent SYM theory, which has boson-fermion degeneracy.  Naively
the absence of the fermionic superpartners of the bosons in the
orbifold theory would suggest that all of the interesting properties
of the supersymmetric spectrum are removed in the orbifold.  However,
supersymmetric theories also have {\it degeneracies among bosons} with
different spin-parity assignments.  For example, scalar and
pseudoscalar fields can be in the same muliplet, as can spin-1 and
spin-0 fields, or spin-2 and spin-1 fields.  Thus the NPO conjecture
implies that the meson spectrum of the $SU(N)\times SU(N)$ theory will
have surprising degeneracies among its bosonic mesons, up to $1/N$
corrections.  Such behavior is quite remarkable, as it is ensured by
no apparent symmetry.

While the spectrum of SYM is poorly understood, there are a few
comments one can make with little difficulty.  First, the phase and
magnitude of the fermion bilinear condensate --- the $\eta'$ and the
$\sigma$ --- are in the same supermultiplet.  They will therefore be
degenerate in the orbifold, for large $N$.  Similar comments govern a
low-lying spin-1 meson and spin-2 glueball, the first of which is the
generalization of the $\omega/\phi$ flavor-singlet vector meson.  It is
possible that additional information can be gleaned by further study
of the SYM spectrum along the lines of \cite{FarGabSch}.  In any case, these
degeneracies will be found throughout the spectrum, along entire Regge
trajectories.  Finally, in addition to degeneracies, we also expect,
more generally, that all hadronic mass ratios should be the same in
the two theories.   

Is it useful to know that the two theories share a hadronic spectrum,
since we do not have predictions for the hadron masses?  Yes --- since
both of these theories can in principle be simulated
\cite{latSQCD,kapschm,flem}.  In fact, they are as easy, and as hard,
to simulate as unquenched QCD with light fermions.  (Actually they
have better infrared behavior than QCD, which has light pions.)  At
least, since these theories have no massless scalars, they do not
suffer from extreme fine-tuning as would most supersymmetric theories
and their orbifolds.  Consequently, the above predictions (along with
the earlier assumption of no phase transition between $g_1>g_2$ and
$g_1=g_2$) may be checked numerically.

Indeed, we can potentially improve the situation for numerical
simulation in two ways.  It is clear that the perturbative
orbifold-parent relation holds even with quenching (although in this
case it becomes relatively trivial.)  We may hope
therefore that both quenched theories are nonperturbatively identical
at large $N$.  More physically, we can avoid the difficulties of
massless fermions by adding equal masses to the gluino, in the SYM,
and to the Dirac fermion in the $\ZZ_2$ orbifold.  While the mass
degeneracies in the spectrum will be lifted, they will be broken in
identical ways in the two theories (at large $N$.)  Also, using
supersymmetry-breaking spurion techniques, it should be possible to
obtain quantitative theoretical predictions for some of the multiplet
splittings when the gluino mass is small.  In any case, the predictions
of the NPO conjecture apply for any mass, so one may start by
comparing the effects of a heavy fermion in the adjoint of $SU(2N)$ to
those of a heavy fermion in the bifundamental of $SU(N)\times SU(N)$,
generally as a function of this mass.

The predictions of the NPO conjecture are by no means limited to the
hadron spectrum.  Let us consider the question of condensates.  If the
$(\ZZ_2)_O$ is preserved in the true vacuum of the orbifold (as was
assumed above) then all condensates are $\ZZ_2$-even and appear in the
supersymmetric parent.  In a supersymmetric theory, no
highest-components of superfields can have condensates without
breaking supersymmetry, which SYM is known not to do.  Therefore,
there are large classes of operators in the orbifold, which descend
from highest componenents of superfields, which will have condensates
that vanish at large $N$.  These include gluon condensates $\vev{\tr
F^2}$, which are forbidden in the supersymmetric theory.  For those
condensates which are nonzero, their magnitudes and phases should be
the same in both orbifold and parent.  The appearance of new
condensates at finite gluino mass should also be mirrored in the
orbifold theory, as should the variation in the fermion bilinear
condensate as a function of the fermion mass.  Again, these effects
could potentially be seen in simulations.

Since with zero fermion mass there are exactly degenerate isolated
vacua (for any $N$) in both theories, there are stable domain walls.
Much is known about these domain walls in the supersymmetric case;
their tensions \cite{dvalshif} and degeneracies \cite{AchVaf} are
exactly predicted.  These exact predictions should also hold in the
orbifold theories for large enough $N$.  

For any value of the fermion mass, these theories have confinement and
corresponding flux tubes charged under $(\ZZ_N)_C$.  While there are
no wholly reliable calculations of the tensions of flux tubes, it
should be possible to compute and compare the flux-tube tensions in
the two theories.\footnote{ Note that the parent $SU(2N)$ has twice as
many flux tubes and domain walls as does the orbifold, which has only
$N$.  This is typical in orbifold projections.  For quantities such
as string tensions, one must be careful to compare the objects which
are supposed to match.}

Even this is not a complete list, but as it will be some years before
these many predictions are tested, let us turn our attention
elsewhere.  What new predictions might be obtained for $\ZZ_p$ orbifolds
with $p>2$?  In this
case no masses can be added for the gluino, since the orbifold
projection removes the mass operator (although supersymmetry-violating
higher-dimensional operators could be added to the theory.)  What
operator would we expect to get a chiral-symmetry breaking expectation
value?  For $p$ odd the lowest-dimension operator charged under
$(\ZZ_{2N})_A$ is the ``double-polygon''
$\tr(\psi_1\psi_2\cdots\psi_p\psi_1\psi_2\dots\psi_p)$; for $p$ even
the lowest operator charged under $(\ZZ_{N})_A$ is the single-polygon
$\tr(\psi_1\psi_2\cdots\psi_p)$.  We would naturally expect these
operators to get expectation values and break the chiral symmetries
down to $\ZZ_2$ for $p$ odd and $U(1)_B$ for $p$ even. 

One can perform a series of checks of this expectation.  Let us take
the couplings away from the orbifold point, so that they are all
different.  Suppose, for example, that $SU(N)_2$ has the largest of
the gauge couplings.  At some scale $\Lambda_2$, $SU(N)_2$ becomes
strongly coupled, with all the other gauge groups acting as
weakly-coupled spectators to its dynamics.  From QCD with three light
flavors, we expect that this $SU(N)$ gauge theory with $N$ quarks
$\psi_2$ and $N$ antiquarks $\psi_1$ will confine and generate a
bilinear fermion condensate $\vev{\psi_1\psi_2}$.  Just as in QCD,
this condensate breaks $SU(N)_1\times SU(N)_3$ (the ``left'' and
``right'' chiral symmetries for the $SU(N)_2$ gauge factor) down to a
diagonal subgroup $SU(N)_D$.  In the breaking, the $N^2-1$ charged
pions of $\psi_1\psi_2$ are eaten by the broken vector bosons (leaving
one neutral pion which decouples from the rest of the theory.)  The
combination of confinement of $SU(N)_2$ and breaking of $SU(N)_1\times
SU(N)_3\rightarrow SU(N)_D$ leaves $p-2$ gauge factors with $p-2$ Weyl
fermions cyclically charged as before.  In short, {\it the $\ZZ_p$
orbifold tumbles dynamically to the $\ZZ_{p-2}$ orbifold.}  The next
stage of confinement and chiral symmetry breaking removes two more
factors, and so on.  This fact was the basis of the ``moose calculus''
of Georgi \cite{georgi} and was used extensively in technicolor model
building.

If $p$ is odd, then the endpoint of this tumbling is simply the $p=1$
case, which is just \none\ SYM.  Thus \cite{kaplan} {\it for odd $p$
the low-energy effective theory has accidental supersymmetry!}  The
final step in the dynamics will involve SYM physics: confinement,
which generates flux tubes with a $(\ZZ_N)_C$ quantum number, and a
breaking of the discrete chiral symmetry, leaving a set of $N$ vacua
which can have domain walls between them.  (Note however this
low-energy theory has a smaller value of $\zeta$ than does the
parent theory, and thus they are not identical; this is not, of
course, a mark against the NPO conjecture, since we have moved far
from the orbifold point to obtain this conclusion.)  Thus when
the couplings are different the qualitative physics expected from
the NPO conjecture is recovered.  If we now assume that there
is no phase transition between the equal-coupling regime and 
the tumbling regime, then the NPO conjecture is quite plausible.
One can go further and check that the tumbling regime predicts
indeed that the double-polygon gets an expectation value, and
moreover that the size of this expectation value is guaranteed
by symmetries to be of the same order as would be expected for
this operator in the parent.

Similar statements can be made for $p$ even, although it is a bit more
intricate.  In this case one can show that the smallest coupling
determines the dynamics in an interesting way.  Suppose $SU(N)_2$ has
the smallest coupling.  Then the {\it odd} gauge factors determine the
size of chiral symmetry breaking condensates, while the {\it even}
ones determine the size of confining string tensions. The reverse
is true if, say, $SU(N)_3$ has the smallest coupling.

Note also, however, that we can view the $(\ZZ_{2q})$ theory as a
$\ZZ_q$ orbifold of the $\ZZ_2$ orbifold itself, or more generally of
the latter theory with unequal gauge couplings.  If $g_1\neq g_2$, we
expect (as discussed earlier) the stronger coupling to determine
chiral condensates and the weaker to determine string tensions.  That
the odd-even structure appears also in the $\ZZ_{2q}$ orbifold adds
some more plausibility to the NPO conjecture.  Again the scales
of the condensates and confinement can be estimated and agree with
expectations in the parent.  Furthermore, there are new classes
of predictions.  For example, consider baryons. The SYM parent has no
baryons, but the $\ZZ_p$ orbifolds all have operators $\det \psi_i$
for each $i$.  Baryons appearing in both the $\ZZ_2$ orbifold and in
the $\ZZ_{2q}$ orbifold should have related spectra, etc.  Of course
it would be much harder to test these relations since the $\ZZ_4$
orbifold is inherently chiral.

 Even with SYM as the parent, there are other
orbifolds than $\ZZ_p$ to consider; the simplest is $SU(2N)\times
SU(N)^4$ with six suitably chosen chiral fermions, a $D_4$ orbifold.
But we need not restrict ourselves to SYM.  Many non-supersymmetric
theories can be orbifolded (indeed we just a moment ago treated the
$\ZZ_2$ orbifold as a parent to $\ZZ_4$) and there are huge classes of
related examples.  For example, one might orbifold $SU(N)\times SU(M)$
with a Dirac fermion to obtain a theory with its own new set of
dynamics\footnote{As this paper was completed, these and other similar
models appeared in a new context, with the $(\ZZ_p)_O$ acting as
translation in a latticized fifth dimension \cite{ACG}.} or add
additional flavors of fermions carrying a nonabelian flavor symmetry
as in QCD.  In these cases we know little about either parent or
orbifold.  However, it may still be very useful to break up the huge
space of nonsupersymmetric theories into a smaller set of classes
which are related nonperturbatively at large, and perhaps not so
large, $N$.  This deserves more study.  Perhaps problems in 0+1 or 1+1
dimensions would give some hints as to how best to do this.

Equally important would be a direct proof (with a clear set of necessary and
sufficient conditions) of the NPO conjecture. 
Presumably one should first study soluble matrix models in
lower dimensions.  There are other more specific questions for which
rigorous answers may be obtainable: does the vacuum of the $\ZZ_p$
orbifold theory necessarily preserve the $(\ZZ_p)_O$ symmetry?  can
one ensure the absence of a phase transition between regimes of
equal and unequal coupling in the $\ZZ_2$ orbifold?  are there other
methods for demonstrating the bosonic hadron degeneracies of the $\ZZ_2$
orbifold?

Most importantly, however, other methods which work in
nonsupersymmetric contexts, perhaps also at large $N$, are needed.
Orbifolding, while potentially a powerful technique, is not enough.
We will need more guidance before the world of nonsupersymmetric
theories becomes transparent.\footnote{As this paper was completed,
the work of \cite{ACG} linked these classes of theories to theories in
latticized extra dimensions.  In their language, the arguments
of this paper imply that at large $N$ the Green's functions
of fifth-dimension zero-modes are independent of the
fifth-dimension lattice spacing.}

\section*{Acknowledgements} I am grateful to many colleagues
for conversations, 
including A. Hanany, V. Kazakov, M. Luty, M. Schmaltz, and
E. Silverstein.  This work was supported by Department of Energy
grant DE-FG02-95ER40893.





\end{document}